# Determining a hopping polaron's bandwidth from its Seebeck coefficient:

# Measuring the disorder energy of a non-crystalline semiconductor

David Emin

Department of Physics and Astronomy

University of New Mexico

Albuquerque, New Mexico 87131 USA

emin@unm.edu

**ABSTRACT**: Charge carriers that execute multi-phonon hopping generally interact strongly enough with phonons to form polarons. A polaron's sluggish motion is linked to slowly shifting atomic displacements that severely reduce the intrinsic width of its transport band. Here a means to estimate hopping polarons' bandwidths from Seebeck-coefficient measurements is described. The magnitudes of semiconductors' Seebeck coefficients are usually quite large ($>k/|q| = 86$ μV/K) near room temperature. However, in accord with the third law of thermodynamics, Seebeck coefficients must vanish at absolute zero. Here the transition of the Seebeck coefficient of hopping polarons to its low-temperature regime is investigated. The temperature and sharpness of this transition depends on the concentration of carriers and on the width of their transport band. This feature provides a means of estimating the width of a polaron's transport band. Since the intrinsic broadening of polaron bands is very small, less than the characteristic phonon energy, the net widths of polaron transport bands in disordered semiconductors approach the energetic disorder experienced by their hopping carriers, their disorder energy.

## I. INTRODUCTION

The absolute Seebeck coefficient $\alpha$ is usually defined as the ratio of the emf generated across a material to the temperature differential that induces it.[1] The Seebeck coefficient is also the entropy transported with a carrier divided by its charge.[1,2]

The entropy transported with a carrier is the sum of (1) the change of a system's entropy upon adding a charge carrier plus (2) the energy associated with moving it divided by the temperature $T$.[2,3] The addition of a charge carrier generally alters (1) the entropy-of-mixing associated with distributing charge carriers among thermally available states, (2) the entropy of atoms' vibrations and (3) the entropy of a material's magnetic moments. Only the carrier-induced change of the system's entropy-of-mixing depends explicitly upon the carrier density. This contribution dominates the Seebeck coefficient except in exceptional circumstances.[2-4]

A semiconductor's carrier density is usually much less than the density of its thermally accessible transport states near room temperature. The magnitude of the semiconductor's Seebeck coefficient then exceeds $k/|q|$ = 86 μV/K, the Boltzmann constant $k$ divided by the magnitude of a carrier's charge $q$. Moreover, the decrease of the Seebeck coefficient's magnitude generated by an Arrhenius increase of its carrier density measures its activation energy.[5-9]

When the ratio of the thermal energy $kT$ to the width of a carrier's transport band $W$ is large enough, all of its states become thermally accessible. The entropy-of-mixing contribution to these carriers' Seebeck coefficient is then given by the Heikes formula: $(k/q)\ \ln[c/(1-c)]$, where $c$ denotes the carrier concentration, the ratio of the densities of carriers to transport states.[10]

The width of a semiconductor's transport band generally becomes exceptionally narrow when its electronic charge carriers self-trap.[11] A self-trapped electronic carrier is bound in the potential well produced by displacements of the atoms that surround it. The composite quasiparticle comprising a self-trapped carrier and the pattern of atomic displacements defines a (strong-coupling) polaron. Such a polaron only moves when the displaced atoms surrounding its self-trapped electronic carrier alter their positions. The intrinsic width of the resulting polaron band is very narrow, much less than the characteristic phonon energy. Furthermore, the polaron's intrinsic bandwidth generally falls as the spatial extent of the polaron's self-trapped carrier is reduced.[11] Thus, the intrinsic bands of severely localized (small) polarons tend to be narrower than those of more extended (large or molecular) strong-coupling polarons.

Materials whose charge carriers form polarons are often disordered. Then the actual width of a band of transported polarons tends to be dominated by the energetic disorder they encounter. Measurements of the Seebeck coefficient can then provide a crude estimate of this energetic disorder.[5,7,9] For example, the observation that the Seebeck coefficients of carriers injected into some organic FETs satisfy the Heikes formula indicates that their transport bands are narrower than $kT$.[12]

Different considerations emerge as the temperature is lowered toward absolute zero. In accord with the third law of thermodynamics, Seebeck coefficients vanish in the limit of zero absolute temperature. Furthermore, the Seebeck coefficients of superconductors remain at zero below their transition temperatures since their ground-state then dominates their transport.

Here the entropy-of-mixing contribution to hopping polarons' Seebeck coefficient is calculated as a function of $c$ and $W$ from high temperatures to very low temperatures. The transition between these two distinct temperature domains provides another estimate of the bandwidth. Since the intrinsic broadening of a polaron band is so small, the width of a transport band in a disordered semiconductor readily rises to be just the spread of its site energies, its disorder energy.

## II. FORMALISM

The change of the entropy-of-mixing arising from the addition of a charge carrier is given by

$$\Delta S = \frac{\partial S}{\partial N_c}\Delta N_c = \frac{\partial[(U-\mu N_c)/T]}{\partial N_c}\Delta N_c = \frac{(E-\mu)}{T}, \quad (1)$$

where $\Delta N_c = 1$. Here the single-particle energy is just the partial derivative of the internal energy $U$ with respect to the carrier number $E \equiv \partial U/\partial N_c$ and the chemical potential associated with the entropy-of-mixing $\mu$ for Fermion carriers is implicitly defined by the relation:

$$N_c \equiv cN \equiv \int dEg(E)f(E,\mu) = \int \frac{dEg(E)}{exp[(E-\mu)/kT]+1}, \quad (2)$$

where $g(E)$ and $f(E,\mu)$ respectively denote the carriers' density-of-states and the Fermi function.

The Seebeck coefficient for a phonon-assisted polaron hop from state $i$ of energy $E_i$ to state $j$ with energy $E_j$ is[2,11,13]

$$\alpha_{i,j} = \frac{E_i\left(\frac{\Gamma_j}{\Gamma_i+\Gamma_j}\right) + E_j\left(\frac{\Gamma_i}{\Gamma_i+\Gamma_j}\right) - \mu}{qT}, \quad (3)$$

where $\Gamma_i$ and $\Gamma_j$ represent the effective electron-phonon coupling strengths at the sites involved in the transition. For jumps between sites with equivalent electron-phonon coupling strength, as expected for most polaron hopping, $\Gamma_i = \Gamma_j$, this Seebeck coefficient becomes simply $\alpha_{i,j} = [(E_i + E_j)/2 - \mu]/qT$.

The net Seebeck coefficient for a system is weighted by its contribution to the electrical conductance. To obtain an expression for the conductance, the net electrical current between states $i$ and $j$ is first written as[11]

$$I_{i,j} = q\left[f_i\left(1 - f_j\right)R_{i,j} - f_j(1 - f_i)R_{j,i}\right]. \quad (4)$$

Now the Fermi factors associated with the probabilities of sites $i$ and $j$ being occupied or unoccupied are expressed in terms of their quasi-electro-chemical potentials $\mu_i$ and $\mu_j$:

$$f_i = \frac{1}{2}exp[-(E_i - \mu_i)/2kT]sech[(E_i - \mu_i)/2kT] \quad (5)$$

and

$$1 - f_j = \frac{1}{2}exp\left[\left(E_j - \mu_j\right)/2kT\right]sech\left[\left(E_j - \mu_j\right)/2kT\right]. \quad (6)$$

Furthermore the requirement of detailed balance between the rates for forward and reverse phonon-assisted hops between sites $i$ and $j$ is utilized to write them as[11,14]

$$R_{i,j} = exp[-(E_j - E_i)/2kT]r_{i,j}(|E_j - E_i|). \quad (7)$$

Upon incorporating these relations into Eq. (4), the expression for the net current between sites *i* and *j* becomes

$$I_{i,j} = \frac{q}{2} sech[(E_i - \mu_i)/2kT] sech[(E_j - \mu_j)/2kT] r_{i,j}(|E_j - E_i|) sinh[(\mu_i - \mu_j)/2kT]. \quad (8)$$

Proceeding to the linear-response regime where the difference between the quasi-electro-chemical potentials is arbitrarily small, the conductance between sites *i* and *j* is written as

$$G_{i,j} = \frac{q^2}{4kT} sech[(E_i - \mu)/2kT] sech[(E_j - \mu)/2kT] r_{i,j}(|E_j - E_i|). \quad (9)$$

By themselves the hyperbolic-secant functions foster transport in states closest to the chemical potential.

The final factor in Eq. (9) restricts a hop's energy disparity. In particular, when $kT$ typically exceeds ~1/3 of the characteristic phonon energy $\hbar\omega$,[11]

$$r_{i,j}(|E_j - E_i|) \cong r_{i,j}(0) exp\left[-(E_j - E_i)^2/8E_b kT\right], \quad (10)$$

and polaron formation requires that its binding energy $E_b >> \hbar\omega$. Thus $r_{i,j}(|E_j - E_i|)$ can be approximated by $r_{i,j}(0)$ except when the disparity between sites' energies is exceptionally large (e.g. $|E_j - E_i| >> (8E_b kT)^{1/2} \cong 0.25$eV with $E_b > 0.3$ eV and $kT = 0.025$ eV). In making this estimate it is recalled that $E_b$ always exceeds at least twice the polaron's jump rate's high-temperature activation energy.[11] Thus the factor $r_{i,j}(|E_j - E_i|)$ does not typically restrict the energy disparity of a high-temperature polaron hop. By contrast, the factor $r_{i,j}(|E_j - E_i|)$ severely restricts a polaron hop's energy disparity in the complementary low-temperature regime, $|E_j - E_i| >> 4E_b \exp(-\hbar\omega/2kT)$:[15]

$$r_{i,j}(|E_j - E_i|) \cong r_{i,j}(0) exp\left[-\frac{|E_j - E_i|}{2kT} + \frac{(E_j - E_i)}{\hbar\omega} - \frac{(E_j - E_i)^2}{4E_b\hbar\omega}\right], \quad (11)$$

where the first term within the exponential's square brackets dominates in the low-temperature limit. Thus, in the low-temperature limit polaron hops are essentially iso-energetic.

Barring major percolation effects the net Seebeck coefficient is approximately the average Seebeck coefficient for an individual hop weighted by its electrical conductance:

$$\alpha \cong \frac{\iint dE_i dE_j g(E_i) g(E_j) G_{i,j}(E_i, E_j) \alpha_{i,j}(E_i, E_j)}{\iint dE_i dE_j g(E_i) g(E_j) G_{i,j}(E_i, E_j)}. \quad (12)$$

At high enough temperatures for $r_{i,j}(|E_j - E_i|)$ to be approximated by $r_{i,j}(0)$ in Eq. (9), Eq. (12) becomes:

$$\alpha \cong \left(\frac{1}{qT}\right) \frac{\int dE g(E) sech[(E-\mu)/2kT](E-\mu)}{\int dE g(E) sech[(E-\mu)/2kT]}. \quad (13)$$

At low enough temperatures for hops to be treated as iso-energetic this Seebeck coefficient becomes

$$\alpha \cong \left(\frac{1}{qT}\right)\frac{\int dE g^2(E) sech^2[(E-\mu)/2kT](E-\mu)}{\int dE g^2(E) sech^2[(E-\mu)/2kT]}. \quad (14)$$

**III. MODEL**

A simple model of the density-of-states is now employed in order to explicitly demonstrate general features of the Seebeck coefficient. With a square band, $g(E) = N/W$ for $-W/2 \le E \le W/2$ the integral involved in the implicit formula for the chemical potential, Eq. (2), is readily evaluated:

$$c = \frac{1}{W}\int_{-W/2}^{W/2} \frac{dE}{exp[(E-\mu)/kT]+1} = \frac{kT}{W} ln\left\{\frac{exp\left[\left(\frac{W}{2}+\mu\right)/kT\right]+1}{exp\left[\left(-\frac{W}{2}+\mu\right)/kT\right]+1}\right\}, \quad (15)$$

where $c \equiv N_c/N$. Solving this equation for the chemical potential yields:

$$\mu(T) = W\left(c-\tfrac{1}{2}\right) - kT ln\left\{\frac{1-exp[-(1-c)W/kT]}{1-exp[-cW/kT]}\right\}. \quad (16)$$

The first contribution to the chemical potential is just its value at absolute zero, $\mu(0)$. The zero-temperature value of the chemical potential is simply the energetic demarcation between filled and empty states of the carriers' energy band. The second contribution to the chemical potential depends on temperature. This temperature dependence is weak for a wide-band metal, $W >> kT$ with c ~ ½. However, this temperature dependence is strong for a narrow-band semiconductor, with small enough $c$ or $(1 - c)$. Above the very low-temperature limit the chemical potential then resides outside of the transport band. As the temperature increases the chemical potential approaches being simply proportional to temperature:

$$\mu(T) - \mu(0) \rightarrow -kT\, ln\left(\frac{1-c}{c}\right). \quad (17)$$

To illustrate this behavior, Fig. 1 plots the $[\mu(T) - \mu(0)]/W$ against $kT/W$ for three values of $c$.

**A. Seebeck coefficient at higher temperatures**

For this model of the density-of-states Eq. (13) becomes:

$$\alpha = 2\left(\frac{k}{q}\right)\frac{\int_{-W/2}^{W/2} dE sech[(E-\mu)/2kT][(E-\mu)/2kT]}{\int_{-W/2}^{W/2} dE sech[(E-\mu)/2kT]}$$

$$= 2\left(\frac{k}{q}\right)\frac{\int_{(-W/2-\mu)/2kT}^{(W/2-\mu)/2kT} dx sech(x) x}{\int_{(-W/2-\mu)/2kT}^{(W/2-\mu)/2kT} dx sech(x)}. \quad (18)$$

Utilizing the expression for the chemical potential, Eq. (16), the upper and lower limits of integration can be written as:

$$\frac{(W/2-\mu)}{2kT} = \frac{W(1-c)}{2kT} + \frac{1}{2} ln\left\{\frac{1 - exp[-(1-c)W/kT]}{1 - exp[-cW/kT]}\right\} \quad (19)$$

and

$$\frac{(-W/2-\mu)}{2kT} = -\frac{Wc}{kT} + \frac{1}{2} ln\left\{\frac{1-exp[-(1-c)W/kT]}{1-exp[-cW/kT]}\right\}. \quad (20)$$

When expanded to second order in *W/kT* these limits become:

$$\frac{(W/2-\mu)}{2kT} = \frac{1}{2}\ln\left[\frac{(1-c)}{c}\right] + \frac{(1-2c)}{48}\left(\frac{W}{kT}\right)^2 + \frac{W}{4kT} \quad (21)$$

and

$$\frac{(-W/2-\mu)}{2kT} = \frac{1}{2}\ln\left[\frac{(1-c)}{c}\right] + \frac{(1-2c)}{48}\left(\frac{W}{kT}\right)^2 - \frac{W}{4kT}. \quad (22)$$

With these limits, Eq. (18) becomes:

$$\alpha \equiv 2\left(\frac{k}{q}\right)\frac{\int_{\frac{1}{2}ln\left[\frac{(1-c)}{c}\right]+\frac{(1-2c)}{48}\left(\frac{W}{kT}\right)^2-\frac{W}{4kT}}^{\frac{1}{2}ln\left[\frac{(1-c)}{c}\right]+\frac{(1-2c)}{48}\left(\frac{W}{kT}\right)^2+\frac{W}{4kT}} dx\, sech(x)x}{\int_{\frac{1}{2}ln\left[\frac{(1-c)}{c}\right]+\frac{(1-2c)}{48}\left(\frac{W}{kT}\right)^2-\frac{W}{4kT}}^{\frac{1}{2}ln\left[\frac{(1-c)}{c}\right]+\frac{(1-2c)}{48}\left(\frac{W}{kT}\right)^2+\frac{W}{4kT}} dx\, sech(x)}. \quad (23)$$

This expression for the Seebeck coefficient manifests its basic asymmetry; it changes sign upon replacing *c* by 1 – *c*.

For a semiconductor's low carrier concentrations [$c$ <<1 or $(1-c)$ << 1], contributions to the integral are limited to large absolute values of $x$. Thus this integral can be evaluated with its hyperbolic secant being replaced by its large-argument approximation.

For definiteness, consider c << 1. The Seebeck coefficient of Eq. (23) then becomes:

$$\alpha \equiv 2\left(\frac{k}{q}\right)\frac{\int_{\frac{1}{2}ln\left[\frac{(1-c)}{c}\right]+\frac{(1-2c)}{48}\left(\frac{W}{kT}\right)^2-\frac{W}{4kT}}^{\frac{1}{2}ln\left[\frac{(1-c)}{c}\right]+\frac{(1-2c)}{48}\left(\frac{W}{kT}\right)^2+\frac{W}{4kT}} dx\, e^{-x}x}{\int_{\frac{1}{2}ln\left[\frac{(1-c)}{c}\right]+\frac{(1-2c)}{48}\left(\frac{W}{kT}\right)^2-\frac{W}{4kT}}^{\frac{1}{2}ln\left[\frac{(1-c)}{c}\right]+\frac{(1-2c)}{48}\left(\frac{W}{kT}\right)^2+\frac{W}{4kT}} dx\, e^{-x}}. \quad (24)$$

Evaluating these simple integrals yields an expression for the Seebeck coefficient to second order in $W/kT$:

$$\alpha = 2\left(\frac{k}{q}\right)\frac{e^{-x}(x+1)\Big|_{\frac{1}{2}ln\left[\frac{(1-c)}{c}\right]+\frac{(1-2c)}{48}\left(\frac{W}{kT}\right)^2-\frac{W}{4kT}}^{\frac{1}{2}ln\left[\frac{(1-c)}{c}\right]+\frac{(1-2c)}{48}\left(\frac{W}{kT}\right)^2+\frac{W}{4kT}}}{e^{-x}\Big|_{\frac{1}{2}ln\left[\frac{(1-c)}{c}\right]+\frac{(1-2c)}{48}\left(\frac{W}{kT}\right)^2-\frac{W}{4kT}}^{\frac{1}{2}ln\left[\frac{(1-c)}{c}\right]+\frac{(1-2c)}{48}\left(\frac{W}{kT}\right)^2+\frac{W}{4kT}}}$$

$$= \left(\frac{k}{q}\right)\left\{ln\left[\frac{(1-c)}{c}\right]+\frac{(1-2c)}{24}\left(\frac{W}{kT}\right)^2+2\left[1-\left(\frac{W}{4kT}\right)coth\left(\frac{W}{4kT}\right)\right]\right\}$$

$$\rightarrow \left(\frac{k}{q}\right)\left\{ln\left[\frac{(1-c)}{c}\right]+\frac{(1-2c)}{24}\left(\frac{W}{kT}\right)^2-\frac{2}{3}\left(\frac{W}{4kT}\right)^2\right\}$$

$$= \left(\frac{k}{q}\right)\left\{ln\left[\frac{(1-c)}{c}\right]-\frac{c}{12}\left(\frac{W}{kT}\right)^2\right\}. \quad (25)$$

An analogous calculation to that for $c << 1$ but for $(1 - c) << 1$ yields

$$\alpha = \left(\frac{k}{q}\right)\left\{ln\left[\frac{(1-c)}{c}\right] + \frac{(1-c)}{12}\left(\frac{W}{kT}\right)^2\right\}. \quad (26)$$

Figure 2 displays the sharp fall-offs of the Seebeck coefficients from their high-temperature limits shifting to lower temperatures with decreasing carrier concentration.

The temperature of the fall-off of the Seebeck coefficient is defined as that for which it falls to half of its high-temperature value. For example, with $c < 1/2$, Eq. (25) indicates that this transition temperature $T_t$ is determined from:

$$\frac{1}{2} ln\left[\frac{(1-c)}{c}\right] = \frac{c}{12}\left(\frac{W}{kT_t}\right)^2, \quad (27)$$

with the transition's sharpness being described by its fractional slope:

$$\left.\frac{\partial ln\alpha}{\partial T}\right|_{T=T_t} = \frac{2}{T_t}. \quad (28)$$

As illustrated in Fig. 2, the transition becomes sharper as its temperature decreases.

**B. Seebeck coefficient at low temperatures**

In the extreme low-temperature limit $[kT << cW, (1 - c)W]$ the chemical potential resides within the transport band at $W(c - ½)$. The low-temperature Seebeck coefficient is then determined by the asymmetry of the position of the chemical potential within the transport band. In particular, for our model the upper- and lower-limits on the integral governing the low-temperature Seebeck coefficient, Eq. (14), become

$$\frac{(W/2-\mu)}{2kT} \rightarrow \frac{W(1-c)}{2kT} \quad (29)$$

and

$$\frac{(-W/2-\mu)}{2kT} \rightarrow -\frac{Wc}{2kT}. \quad (30)$$

With these limits Eq. (14) is readily evaluated:

$$\alpha \cong 2\left(\frac{k}{q}\right)\frac{\int_{-Wc/2kT}^{W(1-c)/2kT} dx\, sech^2(x)x}{\int_{-Wc/2kT}^{W(1-c)/2kT} dx\, sech^2(x)}$$

$$= 2\left(\frac{k}{q}\right)\frac{\int_{-Wc/2kT}^{-\infty} dx\, sech^2(x)x + \int_{-\infty}^{\infty} dx\, sech^2(x)x + \int_{\infty}^{W(1-c)/2kT} dx\, sech^2(x)x}{\int_{-Wc/2kT}^{W(1-c)/2kT} dx\, sech^2(x)}$$

$$= 2\left(\frac{k}{q}\right)\frac{\int_{-Wc/2kT}^{-\infty} dx\, sech^2(x)x + \int_{\infty}^{W(1-c)/2kT} dx\, sech^2(x)x}{\int_{-Wc/2kT}^{W(1-c)/2kT} dx\, sech^2(x)}$$

$$= 2\left(\frac{k}{q}\right)\frac{\int_{Wc/2kT}^{W(1-c)/2kT} dx\, sech^2(x)x}{\int_{-Wc/2kT}^{W(1-c)/2kT} dx\, sech^2(x)} \cong 8\left(\frac{k}{q}\right)\frac{\int_{Wc/2kT}^{W(1-c)/2kT} dx\, e^{-2x}x}{\int_{-\infty}^{\infty} dx\, sech^2(x)}$$

$$= \left(\frac{k}{q}\right)(2x+1)e^{-2x}\Big|_{W(1-c)/2kT}^{Wc/2kT}$$

$$= \left(\frac{k}{q}\right)\left\{\left[\left(\frac{Wc}{kT}+1\right)e^{-Wc/kT}\right] - \left[\left(\frac{W(1-c)}{kT}+1\right)e^{-W(1-c)/kT}\right]\right\}. \quad (31)$$

The steps employed in obtaining the above result are now enumerated. Following the second equality it is noted that the integral between –∞ and ∞ vanishes due to the oddness of its integrand. A simple change of variable is utilized to obtain the expression following the third equality. The resulting two integrals were then approximated by (1) replacing the integrand in the numerator's integral by its value for large argument and (2) extending the integration limits on the denominator's integral from –∞ and ∞. The remaining two integrals are then evaluated to obtain an analytic expression for the low-temperature Seebeck coefficient. As required by symmetry, the Seebeck coefficient for this model reverses sign upon replacing $c$ with $(1 - c)$ and vanishes for the half-filled band, $c = ½$.

Figure 3 plots the low-temperature Seebeck coefficient in units of $k/q$ against $kT/W$ for three values of the carrier concentration. The very small values of the low-temperature Seebeck coefficient are consistent with the higher-temperature curves shown in Fig. 2.

## IV. DISCUSSION

For simplicity this calculation presumed a temperature-independent carrier concentration. A temperature-independent carrier density is sometimes produced with appropriate doping. In addition, the carrier density can be externally controlled when the material is used in a field-effect-transistor (FET). Indeed, Seebeck coefficients have been measured as functions of FETs' injected polaron concentrations.[12] These measurements enable estimates of the temperature of the Seebeck coefficient fall-off, $T_t$. For example, analysis of room-temperature data of Ref. (12) for the organic polymer IDTBT with $c \sim 0.01$ implies that $T_t < 50$ K.

Nonetheless, in many instances most carriers remain bound to dopants. In these situations the transport band's carrier concentration garners a temperature dependence that reflects carriers being thermally liberated from their dopants. For example, above the transition temperature $T_t$ the Seebeck coefficient adopts the form $\alpha \cong (k/q)[E/kT + (1/2)\ln(1/c_0)]$ for $c = c_0^{1/2} \exp(-E/kT)$ when the chemical potential lies between the energy binding carriers to dopants and that of the appropriate edge of the transport band. In accord with the curves of Figs. 2 and 3, the reduction of the transport band's carrier concentration with decreasing temperature increases the Seebeck coefficient and sharpens its eventual fall toward zero. Distinctively, transport at the very lowest temperatures is dominated by unspecified putative states that surround the chemical potential rather than by states within the primary transport band.[2]

In summary, the large carrier-concentration-dependent Seebeck coefficient of a semiconductor drops to zero with decreasing temperature. The temperature and acuteness of this drop-off depend on the carrier concentration and the width of hopping-carriers' transport band. Thus measurement of the Seebeck coefficient through this transition provides a means of estimating the width of the transport band. In a sufficiently disordered polaronic semiconductor the width of a polaron band is primarily determined by the energetic disorder experienced by its charge carriers. Then observation of the fall-off of its Seebeck coefficient with decreasing temperature provides a means of assessing this energetic disorder.

# Figure Captions

Fig. 1 The difference between the chemical potential and its zero-temperature value in units of the transport band's width $W$, $[\mu(T) - \mu(0)]/W$, is plotted against $kT/W$ for carrier concentrations of $c$ = 0.001, 0.005 and 0.009. With decreasing carrier concentration and rising temperature the chemical potential approaches proportionality to temperature as it falls outside of the transport band.

Fig. 2 The entropy-of-mixing contribution to the Seebeck coefficient $\alpha$ in units of $k/q$ is plotted against $kT/W$, the thermal energy $kT$ in units of the transport band's width $W$, for carrier concentrations of $c$ = 0.001, 0.005 and 0.009. The drop-off of the Seebeck coefficient shifts to lower temperatures with decreasing carrier concentration.

Fig. 3 The entropy-of-mixing contribution to the Seebeck coefficient $\alpha$ in units of $k/q$ is plotted against $kT/W$ in the very low temperature regime for carrier concentrations of $c$ = 0.001, 0.005 and 0.009. The semiconductor's Seebeck coefficient collapse toward zero sharpens with decreasing carrier concentration.

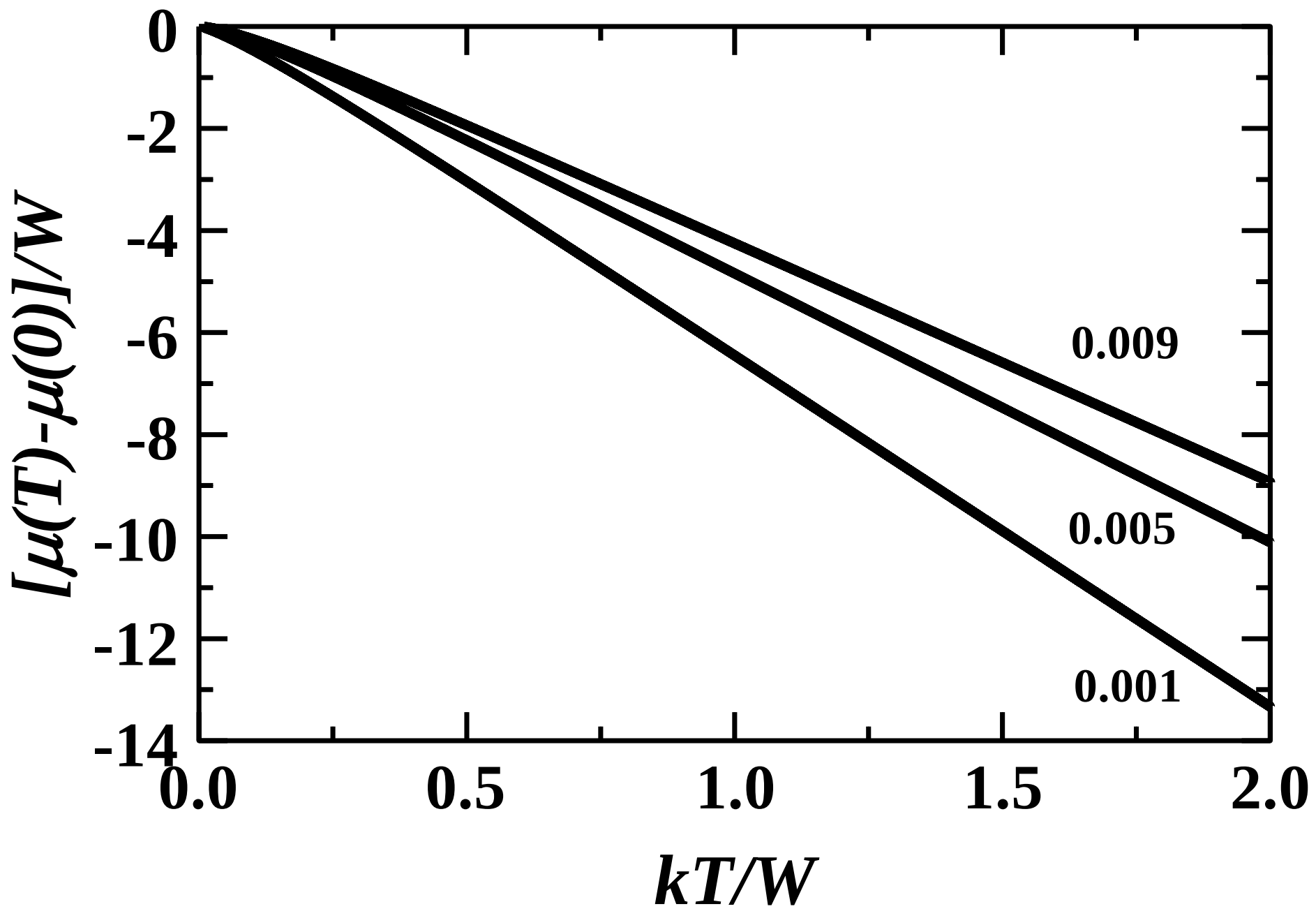

Fig. 1 The difference between the chemical potential and its zero-temperature value in units of the transport band's width $W$, $[\mu(T) - \mu(0)]/W$, is plotted against $kT/W$ for carrier concentrations of $c = 0.001$, 0.005 and 0.009. With decreasing carrier concentration and rising temperature the chemical potential approaches proportionality to temperature as it falls outside of the transport band.

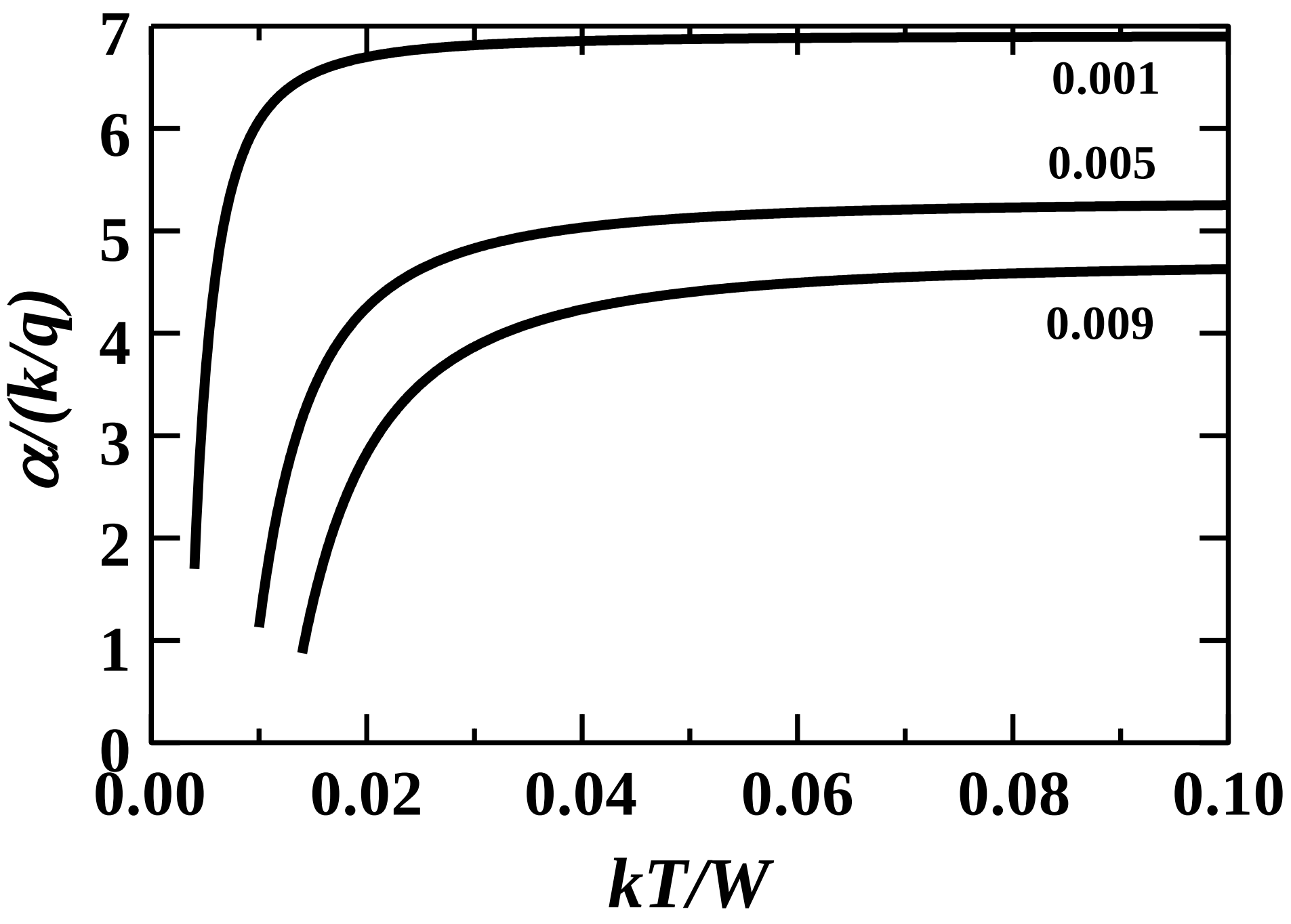

Fig. 2 The entropy-of-mixing contribution to the Seebeck coefficient $\alpha$ in units of $k/q$ is plotted against $kT/W$, the thermal energy $kT$ in units of the transport band's width $W$, for carrier concentrations of $c$ = 0.001, 0.005 and 0.009. The drop-off of the Seebeck coefficient shifts to lower temperatures with decreasing carrier concentration.

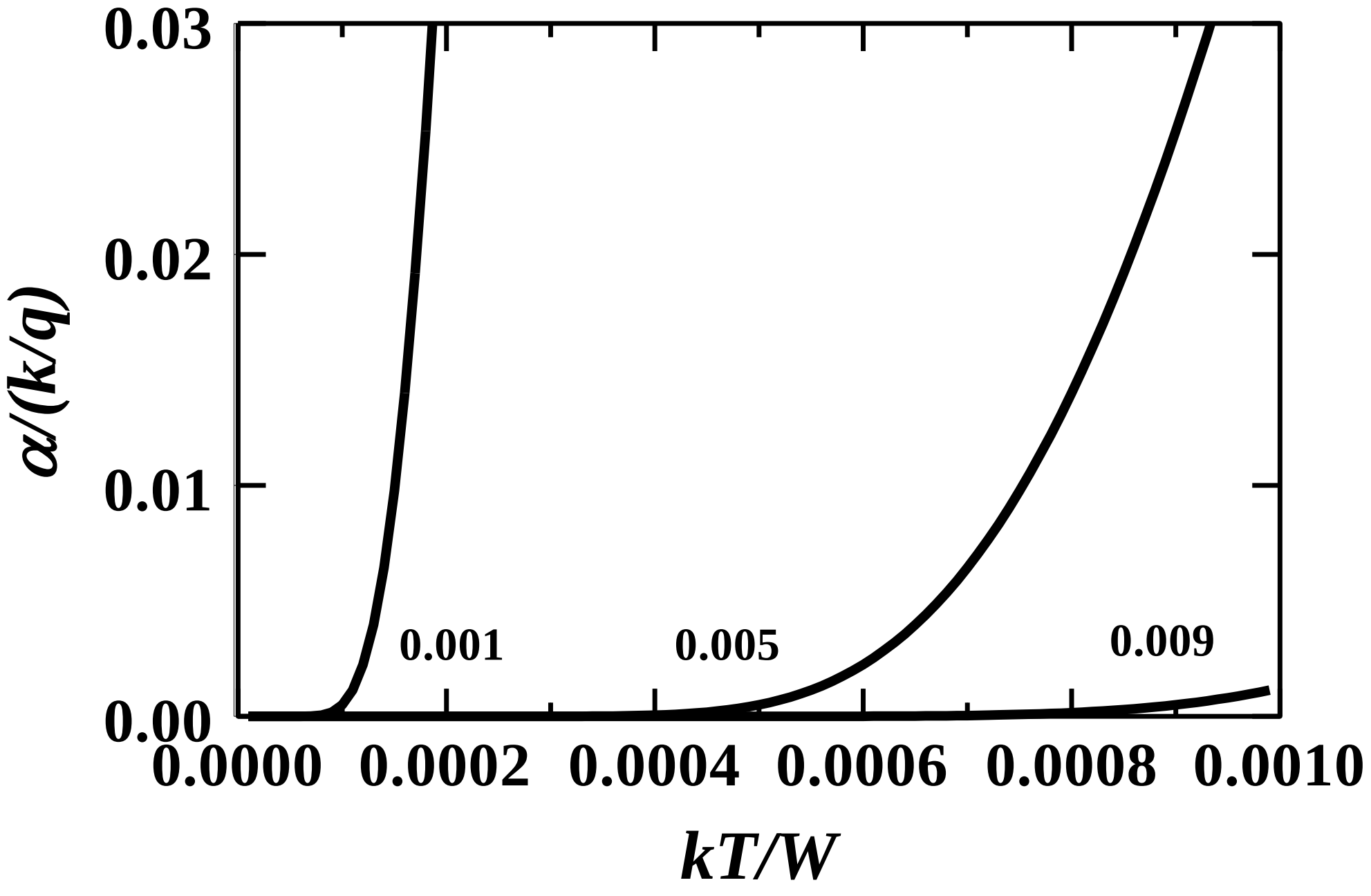

Fig. 3 The entropy-of-mixing contribution to the Seebeck coefficient $\alpha$ in units of $k/q$ is plotted against $kT/W$ in the very low temperature regime for carrier concentrations of $c$ = 0.001, 0.005 and 0.009. The semiconductor's Seebeck coefficient collapse toward zero sharpens with decreasing carrier concentration.